\documentclass[acp]{copernicus}
\usepackage{graphicx}
\begin{document}

\title{\bf Extreme Associated Functions: 
Optimally Linking Local Extremes to Large-scale Atmospheric
Circulation Structures} 

\author{Debabrata Panja} 
\author{Frank M. Selten}
\affil{Royal Netherlands Meteorological
Institute (KNMI), Postbus 201, 3730 AE De Bilt, The Netherlands} 
\runningauthor{Debabrata Panja}
\runningtitle{Linking Local Extremes to Atmospheric
Circulations}
\correspondence{D. Panja (dpanja@science.uva.nl)}
\maketitle
\begin{abstract}
We present a new statistical method to optimally link local weather
extremes to large-scale atmospheric circulation structures. The method
is illustrated using July-August daily mean temperature at 2m height
(T2m) time-series over the Netherlands and 500 hPa geopotential height
(Z500) time-series over the Euroatlantic region of the ECMWF reanalysis
dataset (ERA40). The method identifies patterns in the Z500
time-series that optimally describe, in a precise mathematical
sense, the relationship with local warm extremes in the
Netherlands. Two patterns are identified; the most important one
corresponds to a blocking high pressure system leading to subsidence
and calm, dry and sunny conditions over the Netherlands. The second
one corresponds to a rare, easterly flow regime bringing warm, dry air
into the region. The patterns are robust; they are also identified in
shorter subsamples of the total dataset. The method is generally applicable and
might prove useful in evaluating the performance of climate models in
simulating local weather extremes.
\end{abstract}

\introduction\label{sec1}

Weather extremes such as extreme wind speeds, extreme precipitation or
extreme warm or cold conditions are experienced locally. They are
usually connected to circulation structures of much larger scale in
the atmosphere. For example, if we restrict ourselves to the
Netherlands, a well-known circulation structure that often leads to
extreme hot summer days is a high pressure system that blocks the
inflow of cooler maritime air masses. Moreover, the subsidence of air
in its interior leads to clear skies and an abundance of sunshine that
leads to high temperatures. If the blocking high persists and depletes
the soil moisture due to lack of precipitation and increased
evaporation, temperatures tend to soar, as it did in the European
summer of 2003 \cite{scharetal}. Speculations about a positive
feedback of dry soil on the persistence of the blocking high can also
be found in the literature \cite{ferranti}.
\begin{figure*}[htpb]
\begin{center}
\begin{minipage}{0.48\linewidth}
\includegraphics[width=0.64\linewidth,angle=270]{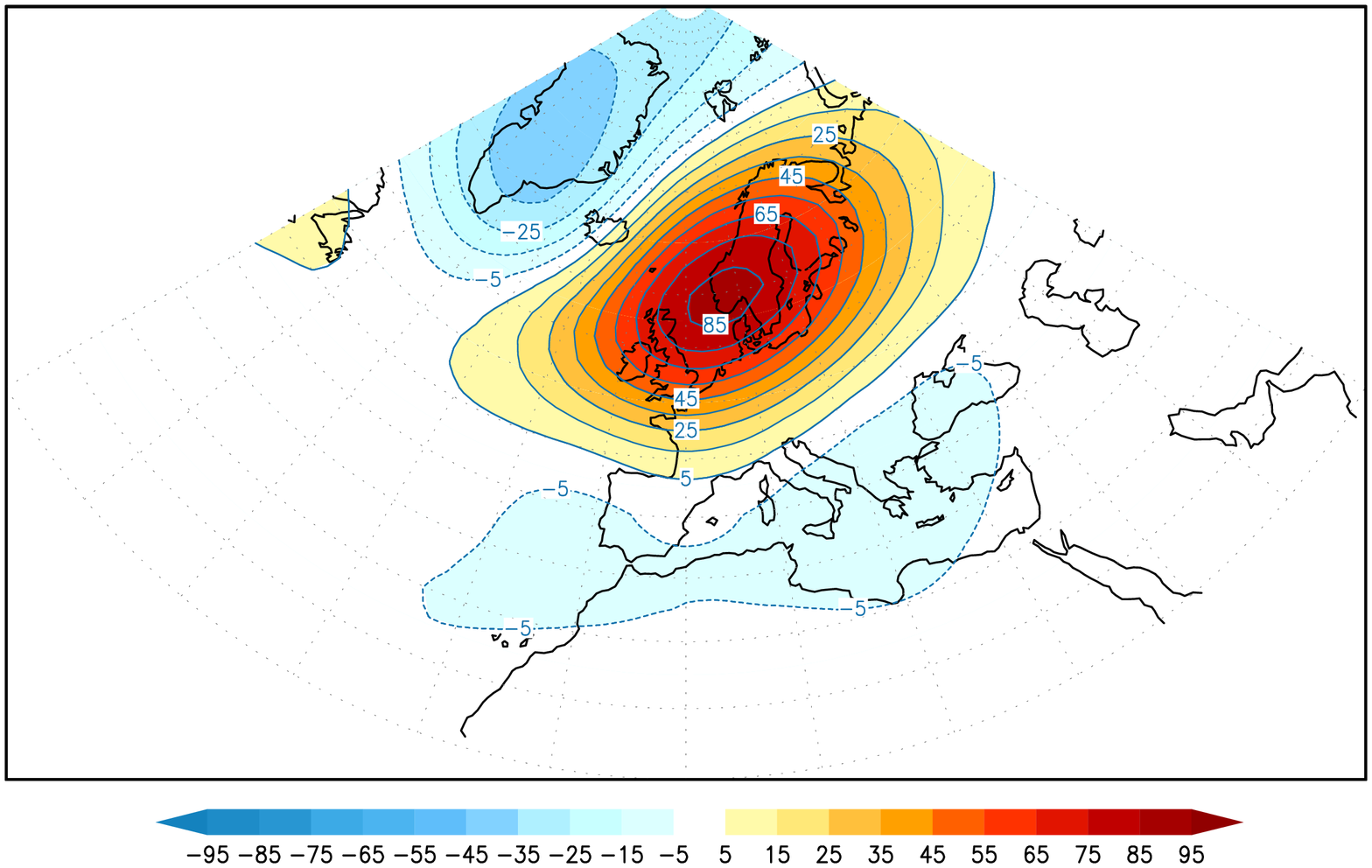}
\end{minipage}
\hspace{2mm}
\begin{minipage}{0.48\linewidth}
\includegraphics[width=0.64\linewidth,angle=270]{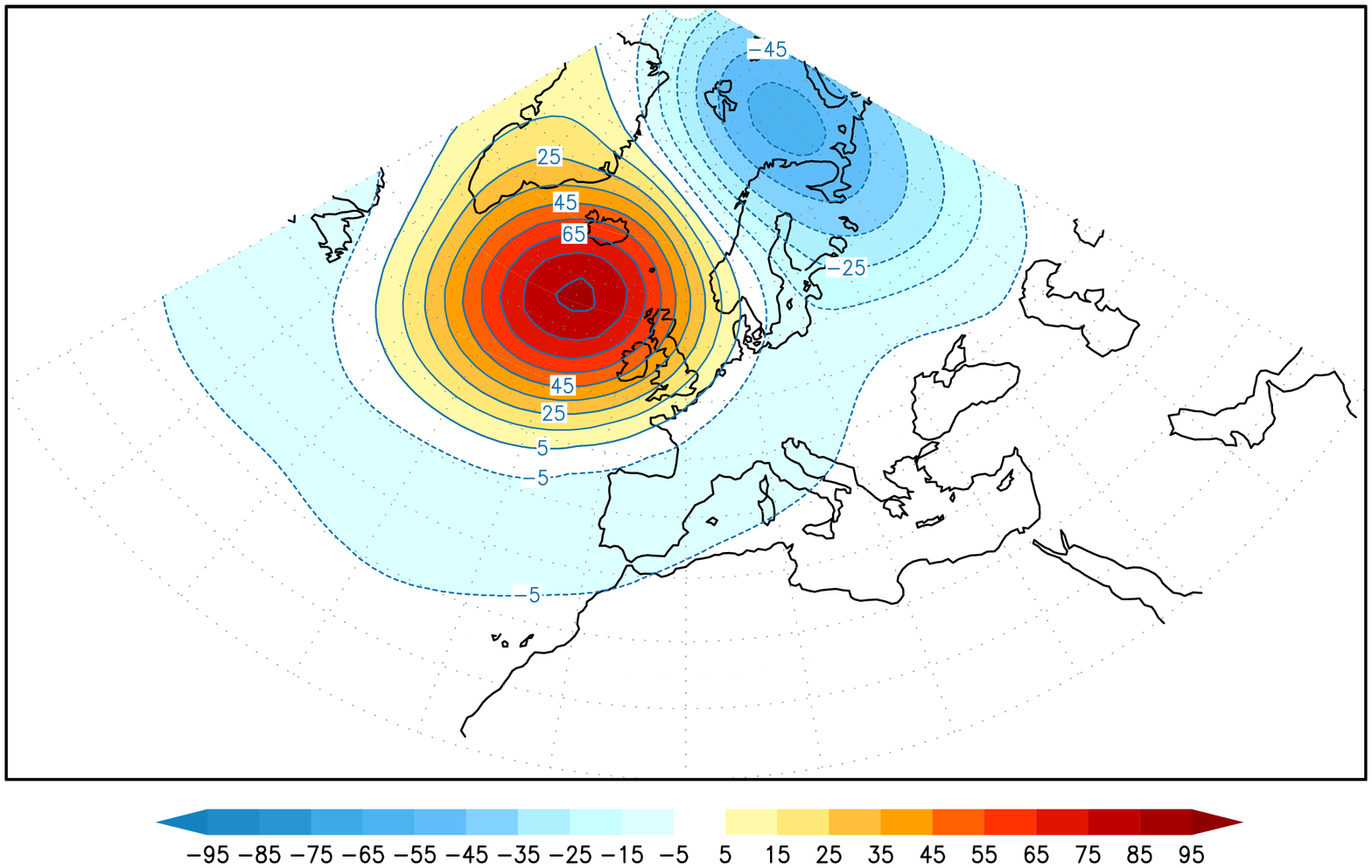}
\end{minipage}
\end{center}
\caption{The leading two EOFs for the July and August Z500 daily
anomaly field for 43 years of ERA-40 data (1958-2000). Left figure:
first EOF; right figure: second EOF. Relative importances are
$12.57\%$ and $11.79\%$ respectively. The patterns have been multiplied by
 one standard deviation of the corresponding amplitude
time-series (in meters).}
\label{fig1}
\end{figure*}

In order for climate models to correctly simulate the probability
of extreme hot summer days, a crucial ingredient is the correct
simulation of the probability of the occurrence of blocking. This is a
well-known difficult feature of the atmospheric circulation to
simulate realistically \cite{pelly}. The verification of
models w.r.t. this aspect is, in practice, difficult as well, since
idealized model experiments suggest a high degree of internal
variability of blocking frequencies even on decadal timescales
\cite{liu}.

In a world with increasing concentrations of greenhouse gases, not
only the temperature increases, also the large-scale circulation
adjusts to achieve a new (thermo)dynamical balance. Models disagree on
the magnitude and even the direction of this change locally
\cite{ulden}. For instance, a change in the probability of European
blocking conditions in summer immediately impacts the future
probability of European heat waves. This makes probability estimates
of future European heat waves very uncertain.  To address the
questions concerning the probability of future extreme weather events,
and the evaluation of climate model simulations in this respect, it is
necessary to have a descriptive method that links local weather
extremes to large-scale circulation features. To the best of our
knowledge, an optimal method to do so does not exist in the literature.

We identified two approaches in the literature to link local weather
extremes to large-scale circulation features. In the first one, the
circulation anomalies are classified first, the connection with local
extremes is analyzed in second instance. The ``Grosswetterlagen''
developed by synoptic meteorologists for instance is one such
classification \cite{kysely}. All kinds of clustering algorithms are
another example \cite{plaut,cassou}. In our opinion, this
approach is not optimal since in the definition of the patterns,
information about the extreme is not taken into account.

In the second approach, a measure of the local extreme does enter the
definition of the large-scale circulation patterns. For instance, only
atmospheric states are considered for which the local extreme occurs.
Next a simple averaging operator is applied [``composite method" as in
\cite{schaef}]or a clustering analysis is performed \cite{sanchez} .
The composite method falls short since it finds by definition only
one typical circulation anomaly and from synoptic experience we know
that often different kind of circulation anomalies lead to a similar
local weather extreme. The clustering analysis is debatable since the
data record is often too short to identify clusters with enough
statistical confidence \cite{hsu}.

The purpose of this paper is to report a new {\it optimal \/} method
to relate local weather extremes to characteristic circulation
patterns. This method objectively identifies, in a robust manner, the
different circulation patterns  that favor the occurrence of local
weather extremes. The method is inspired by the Optimal Autocorrelation
Functions of \cite{selten}. It is based on considering linear combinations
of the dominant Empirical Orthogonal Functions that maximize a suitable
statistical quantity. We illustrate our method by analyzing the
statistical relation between extreme high daily mean temperatures at
two meter height (T2m) in July and August in the Netherlands and the
structure of the large-scale circulation as measured by the 500 hPa
geopotential height field (Z500). 

This paper is divided into five sections. Section \ref{sec2} is
focused on the data, where we explain the method to obtain the daily
Z500 and T2m anomalies in Europe, and report the results of the EOF
analysis of the Z500 anomaly data. In Sec. \ref{sec3} we outline the
procedure to optimize the quantity that describes the statistical
relation between the Z500 and the extreme T2m anomalies, supported by
the additional details in the Appendix. In Sec. \ref{sec4} we
identify the large-scale Z500 anomaly patterns that are associated
with hot summer days in the Netherlands, demonstrate the
robustness of our method and compare the patterns with patterns
earlier reported in the literature. Finally we conclude this report in
Sec.  \ref{sec5} with a discussion on the possible applications of our
method.

\section{The T2m and Z500 datasets, and EOF analysis of the Z500 data
  \label{sec2}} 

Our data have been obtained from the ERA-40 reanalysis dataset, for the
timespan Sept. 1957 to Aug. 2002, at 6 hourly intervals on a
$2.5^\circ\times2.5^\circ$ latitude-longitude grid. These data are
publicly available at the ECMWF website
$\mathsf{http://data.ecmwf.int/data/d/era40\_daily/}$. The T2m data
over entire Europe, defined by $37.5^\circ$N-$70^\circ$N and
$10^\circ$W-$40^\circ$E, and the Z500 data over
$20^\circ$N-$90^\circ$N and $60^\circ$W-$60^\circ$E were
downloaded. From these, the daily averages for T2m and Z500 fields for
the years 1958-2000 (all together 43 years in total) were
computed. This formed our full raw dataset.

In order to remove possible effects of global warming in the last
decades of 20th century, detrending these fields prior to performing
further calculations would be necessary. However, an analysis
of the Z500 daily averaged field revealed no significant linear trend over
these 43 years. Therefore the Z500 daily
anomaly field was  obtained by simply removing the seasonal cycle
defined by an average over the entire period of 43 years. Greatbatch
and Rong (2006) showed that over Europe, the trends in the
ERA-40 reanalysis and NCEP-NCAR reanalysis are indeed small and
similar.

A warming trend, however, is clearly present in the T2m field. For
detrending the T2m field, the monthly averages for July and August
were calculated from the daily averages at each gridpoint. Next,
11-year running means were computed for these monthly averaged  T2m
fields (for July and August separately), and that formed our baseline
for calculating daily T2m anomaly field. This procedure does not yield
the baseline for the first and the last 5 years (1958-1962 and
1996-2000); these were computed by extrapolating the baseline trend
for the years 1963-1964 and 1995-1996 respectively.

For the EOF analysis of the Z500 anomaly field, note that most of the
variance of atmospheric variability resides in the low-frequency part
[10-90 day range \cite{malone}]. Indeed, the dominant EOFs of Z500
anomaly fields proved insensitive to the application of 3-day, 5-day,
7-day, 9-day and 15-day running mean filters. For the sake of
simplicity, therefore, we decided to only consider EOFs based on daily
Z500 anomaly fields. The EOF analysis was performed on the regular
lat-lon grid data with each grid point weighted by the cosine of its
latitude to account for the different sizes of the grid cells. Using
these weights, the EOFs $\mathbf{e}_k$ are orthogonal in space (note
here that we use the same definition of vector dot product in space
all throughout this paper)
\begin{eqnarray}
\mathbf{e}_k \cdot \mathbf{e}_l &\equiv& \nonumber \\&& 
\hspace{-15mm}\frac{1}{\sum_{i=1}^{N}\cos(\phi_i)} \sum_{i=1}^{N} e_k(\lambda_i,\phi_i)
e_l(\lambda_i,\phi_i)\cos(\phi_i)=\delta_{kl}\,,
\label{eq:dot}
\end{eqnarray}
where $\phi$ denotes latitude, $\lambda$ longitude and $N$ the total
number of grid points, and $\delta_{kl}$ is the Kronecker delta. Each
Z500 anomaly field can be expressed in terms of the EOFs as
\begin{eqnarray}
\mathbf{Z500}(t) = \sum_{k=1}^{N} a_k(t)\mathbf{e}_k\,,
\label{eq:expansion}
\end{eqnarray}
where the amplitudes $a_k$ are found by a projection of the Z500
anomaly on to the EOFs
\begin{eqnarray}
a_k(t) = \mathbf{Z500}(t) \cdot \mathbf{e}_k\,.
\label{eq:a}
\end{eqnarray}
A nice property of the EOFs is that their amplitude time-series are
uncorrelated in time at zero lag
\begin{eqnarray}
\left< a_k(t) a_l(t) \right> = \sigma_k^2\, \delta_{kl\,},
\label{eq:uncorrel}
\end{eqnarray}
where the angular brackets $\langle .\rangle$ denote a time average and
$\sigma_k^2$ denotes the eigenvalue of the $k$-th EOF which is equal
to the variance of the corresponding amplitude time-series.

We found that July and August months produced very similar EOFs, while June and
September EOFs were significantly different. We therefore decided to
restrict the summer months to July and August. The leading two EOFs
for the corresponding daily Z500 anomaly for 1958-2000 are shown in
Fig. \ref{fig1}. The values correspond to one standard deviation of
the corresponding amplitude. The two EOFs are not well
separated (the eigenvalues are close together) and therefore we expect
some mixing between the two patterns \cite{north}. A linear
combination of the two EOFs shifts the longitudinal position of the
strong anomaly over Southern Scandinavia which is present in the first
EOF. It resembles the summer NAO pattern as diagnosed by Greatbatch
and Rong \cite{greatbatch} (their figure 8).

\section{Optimization procedure to establish the connection between
Z500 anomalies and local extreme T2m\label{sec3}}

\begin{figure}[!h]
\begin{center}
\includegraphics[width=0.78\linewidth,angle=270]{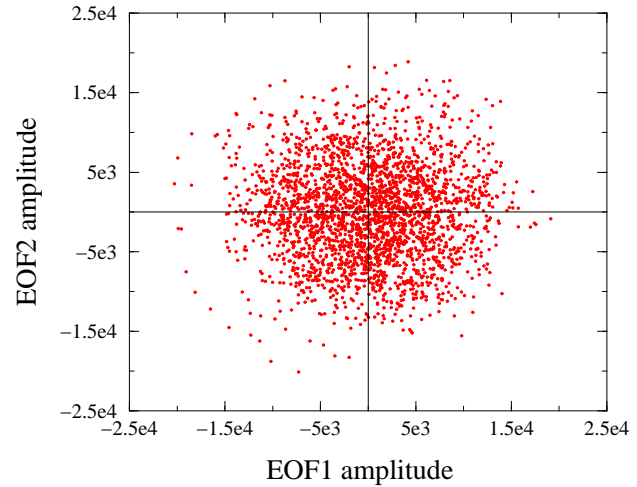}
\end{center}
\caption{Projection of the daily Z500 anomaly field for July and
August months for 43 years in the two-dimensional vector space spanned by
the two leading EOFs.}
\label{fig2}
\end{figure}

One of the first approaches we considered to establish the connection
between Z500 daily anomaly fields and extreme daily T2m is the
so-called ``clustering method", which identifies clusters of points in
the vector space spanned by the dominant EOFs. The daily Z500 anomaly
field for July and August over 43 years yields us precisely 2666
datapoints in this vector space. A projection of these daily anomalies
on the two-dimensional vector space of the two leading EOFs (EOF1 and
EOF2) is shown in Fig. \ref{fig2}. No clear clusters are apparent by
simple visual inspection. One can imagine that defining clusters using
existing cluster algorithms to identify clusters of points that
correspond to specific large-scale circulation patterns that occur
significantly more frequently than others is not a trivial
undertaking. Often it turns out that using 40 years of data or so, the
clusters identified  are the result of sampling errors, due to too few
data points \cite{hsu,berner,stephenson}.

Nevertheless, when we plot the T2m positive anomaly values at the
center of the Netherlands ($52.5^\circ$N, $5^\circ$E) in a scatter
plot with the amplitude of EOF1, a distinct ``tilt" in the scatter
plot emerges: i.e., with increasing amplitude of the leading EOF, the
likelihood of having very hot summer days increases. Having inspected
the same plots for the other EOFs we found a similar tilt for some of
the other EOFs as well. From this point of view, finding the
statistical relationship between T2m at a given place and the state
of the large-scale atmospheric circulation can be reduced to a
mathematical exercise that finds those linear combinations of EOFs
that optimally bring out this tilt. In the remainder of this section,
supported by the Appendix, we present a {\it general, rigorous and
robust\/} procedure to achieve this.
\begin{figure}[!h]
\begin{center}
\includegraphics[width=0.82\linewidth,angle=270]{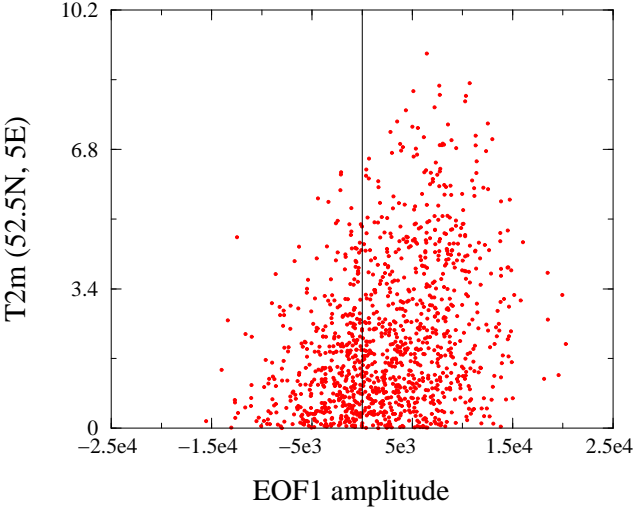}
\end{center}
\caption{Scatter plot of T2m $\geq0$ at the center of the Netherlands vs. the
amplitude of leading Z500 EOF (EOF1). With increasing amplitude of the leading
EOF, the likelihood of having very hot summer days increases.}
\label{fig3}
\end{figure}
\begin{figure*}
\begin{center}
\begin{minipage}{0.49\linewidth}
\includegraphics[width=0.78\linewidth,angle=270]{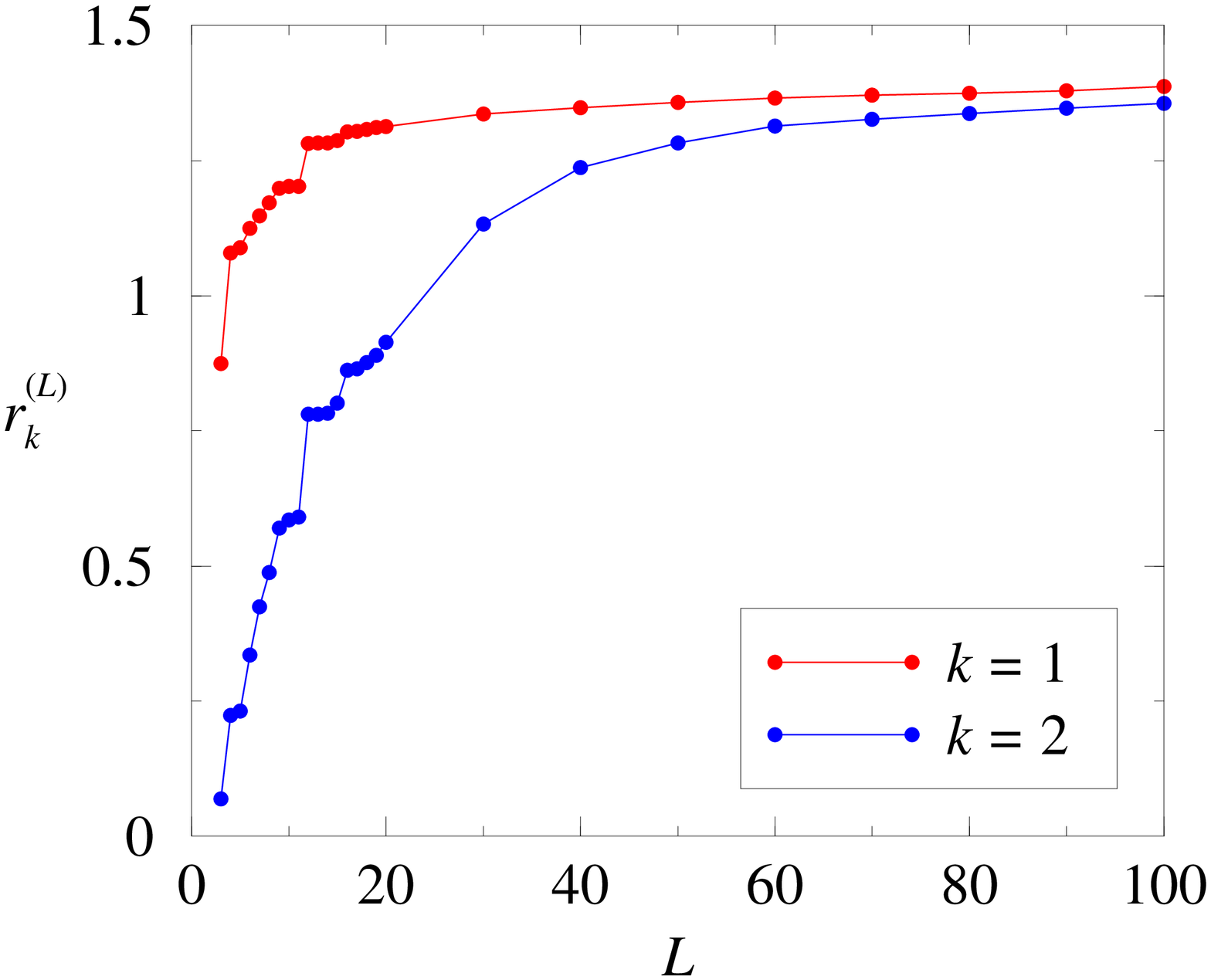}
\end{minipage}
\begin{minipage}{0.49\linewidth}
\includegraphics[width=0.78\linewidth,angle=270]{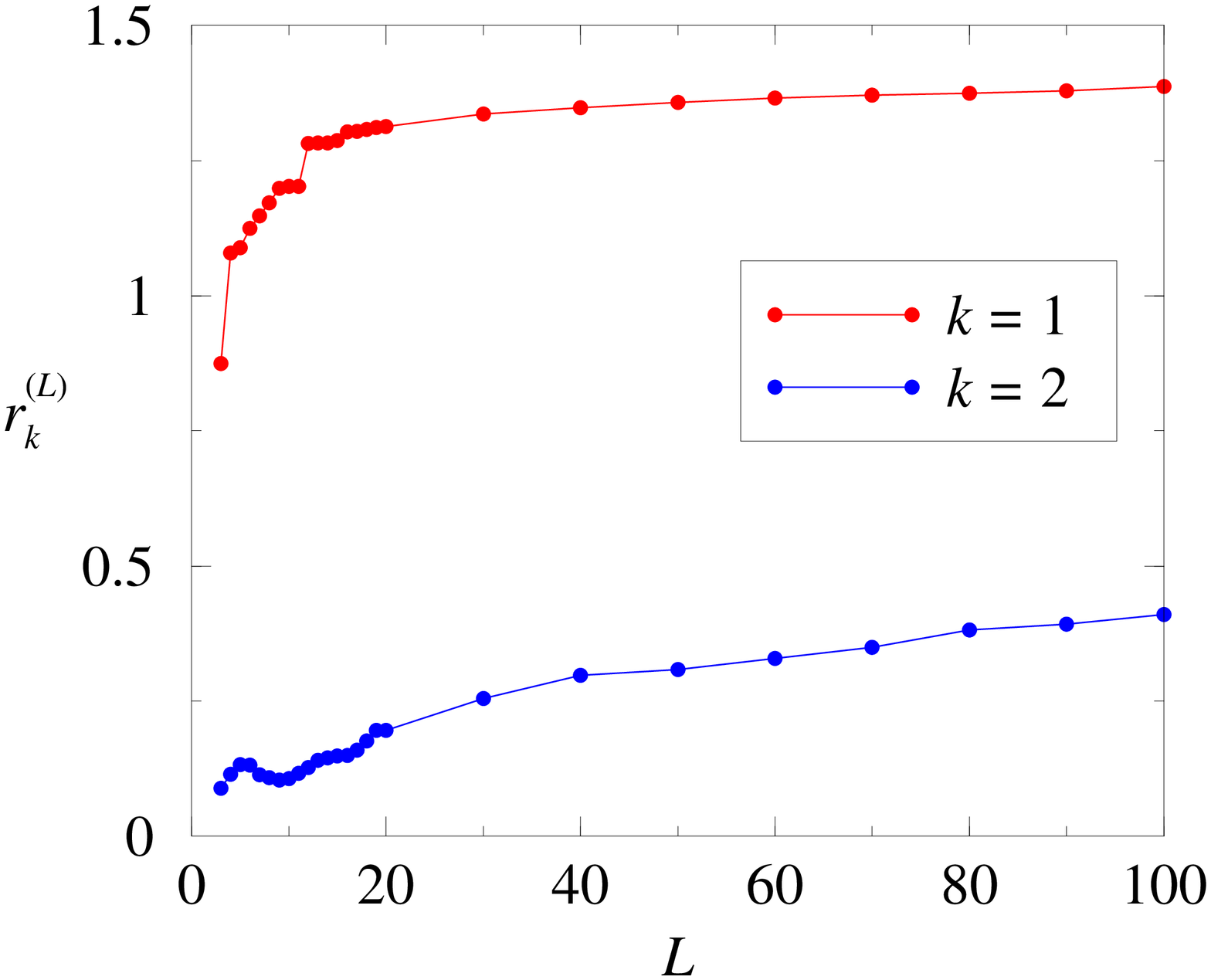}
\end{minipage}
\end{center}
\caption{The behavior of $r$  as a function of $L$ for the first
(red) and second (blue) EAFs. Left panel: patterns are orthogonal in
space, but are correlated in time [option (a) in text]; right panel:
patterns are uncorrelated in time but are not orthogonal in space
[option (b) in text]. }
\label{fig4}
\end{figure*}

To represent this statistical relationship, we start by defining the
following dimensionless quantity
\begin{eqnarray}
r_{k}^{(L)}=\frac{\langle b_{k}^{(L)}(t)\,[T(t)]^n\rangle_{\mbox{\scriptsize p}}}{\langle
\left[b_{k}^{(L)}(t)\right]^2\rangle^{\frac{1}{2}}_{\mbox{\scriptsize p}}\,\,\langle
[T(t)]^{n}\rangle_{\mbox{\scriptsize p}} }.
\label{eq:r}
\end{eqnarray}
Here the angular brackets $\langle .\rangle_{\mbox{\scriptsize p}}$ denotes a time
average taken only over those days for which T2m$(t)\geq0$, and $n$
is a positive number $>1$. The idea behind choosing $n>1$ is that for
higher $T(t)$ it gives larger contribution to $r_{k}^{(L)}$:
we are interested in high-temperature days at gridpoint $G$, we
choose $n=2$ for this study. The variable $b_{k}^{(L)}(t)$ is the
amplitude on day $t$ of a pattern, defined as a linear combination
of the first $L$ EOFs. Since $L$ linear combinations can be defined that
form a new complete basis in the subspace of the first $L$ EOFs we use the
subscript $k$ to denote these different linear combinations. 

We first  concentrate on the calculation of the first pattern.  Using
$c^{(L)}_{j1}$ to denote the coefficients of this first linear
combination then
\begin{eqnarray}
b_{1}^{(L)}(t)=\sum_{j=1}^{L}c^{(L)}_{j1}\,a_j(t)\,.
\label{eq:b}
\end{eqnarray}

Notice that since the time averages are
taken only over those days for which T2m$(t)\geq0$, $\langle
b_{1}^{(L)}(t)\rangle_{\mbox{\scriptsize p}}\neq0$, although $\langle
b_{1}^{(L)}(t)\rangle=0$ since $\langle
a_j(t)\rangle=0$. 

Equations (\ref{eq:r}-\ref{eq:b}) imply that given the time-series of
T2m and Z500 anomalies, the numerical value of $r^{(L)}_1$ depends only on
$L$ and on the coefficients $c^{(L)}_{j1}$. For a given value of $L$,
$c^{(L)}_{j1}$ are found by maximizing the square of $r^{(L)}_1$
within the vector space of the first $L$ EOFs (the square is taken
since $r^{(L)}_{1}$ can  take on negative values as well).

If we define
 for $T(t)\geq0$,  
\begin{eqnarray}
\tilde{T}_{\mbox{\scriptsize p}}(t)=\frac{[T(t)]^{n}}{\langle[T(t)]^{n}\rangle_{\mbox{\scriptsize p}}}
\label{tptilde}
\end{eqnarray}
then Eq. (\ref{eq:r}) can be rewritten for $k=1$ as
\begin{eqnarray}
r^{(L)}_{1}=\frac{\langle b^{(L)}_{1}(t)\,\tilde{T}_{\mbox{\scriptsize p}}(t)\rangle_{\mbox{\scriptsize p}}}
{\langle \left[b^{(L)}_{1}(t)\right]^2\rangle^{\frac{1}{2}}_{\mbox{\scriptsize p}}}.
\label{eq:ralternative}
\end{eqnarray}
In words, maximizing $\left[r^{(L)}_1\right]^2$ defines a pattern that
for a change of one standard deviation in its amplitude $b_1$ brings
about the largest change in the normalized  positive temperature
anomaly $\tilde{T}_{\mbox{\scriptsize p}}$ or put differently the local
temperature responds  most sensitively to changes in the normalized
amplitude of this pattern. In this sense, the pattern is  optimally
linked to the local warm temperature extremes.

It is shown in the appendix that maximizing $\left[r^{(L)}_1\right]^2$
corresponds to the linear least squares fit  of the EOF amplitude
timeseries to $\tilde{T}_{\mbox{\scriptsize p}}(t)$
\begin{eqnarray}
\tilde{T}^{(L)}_{\mbox{\scriptsize p}}(t) = \sum_{j=1}^{L}r^{(L)}_{1}c^{(L)}_{j1}\,a_j(t).
\label{eq:regression}
\end{eqnarray}
with the coefficients $c_{j1}$ given by
\begin{eqnarray}
r^{(L)}_{1}c^{(L)}_{j1} = \sum_{i=1}^{L} \langle a_i(t)a_j(t)\rangle_{\mbox{\scriptsize p}}^{-1}\,\langle 
\tilde{T}_{\mbox{\scriptsize p}}(t)\,a_i(t) \rangle_{\mbox{\scriptsize p}}
\label{eq:rcoeff}
\end{eqnarray} 

This result makes sense since the linear least squares fit optimally
combines the  EOF amplitude timeseries to minimize the mean squared
error between the actual  temperature anomaly and the temperature
anomaly estimated from the circulation  anomaly at that day.

The procedure to find the remaining $(L-1)$ linear combinations is as
follows. We  first reduce the $Z500$ anomaly fields to the $(L-1)$
dimensional subspace $\mathbf{Z500}^{(L-1)}$  that is orthogonal to
the first linear combination. In this subspace we again determine  the
linear combination that optimizes $r^{(L)}_2$. By construction, this
value is lower than $r^{(L)}_1$. This procedure is repeated to
determine all $L$ linear combinations with  decreasing order of
optimized values $r^{(L)}_k$.

There is no unique way to define the subspaces and how this is done
affects the  properties of the linear combinations.  The linear
combinations can either (a) be constructed to form an  orthonormal
basis in space, in which case their amplitudes are temporally
correlated;  or (b) they can be constructed so that the corresponding
amplitudes are temporally uncorrelated, but  in that case they are not
orthonormal in space.  In both cases, they form a complete basis in
the space of the first $L$ EOFs
\begin{eqnarray}
\mathbf{Z500}^{(L)}(t)=\sum_{k=1}^{L} b^{(L)}_k(t)\,\mathbf{f}^{(L)}_k\,.
\label{eq:expansionf}
\end{eqnarray}

We will call the patterns $\mathbf{f}^{(L)}_k$ Extreme Associated
Functions (EAFs). The mathematical details on how to obtain
$b^{(L)}_k(t)$ for both options can be found in the Appendix.

\section{Statistical relationship between high summer
temperature in the Netherlands and large-scale atmospheric
circulation structures\label{sec4}} 

We now need a criterion to determine the optimal number of EOFs in
the linear combinations. The reason for limiting the number of EOFs in
the linear combinations is apparent from Eq. (\ref{eq:rcoeff}). Here
the inverse of the covariance matrix of the EOF amplitudes
appears. This matrix becomes close to singular when low-variance EOFs
are included in the linear combination. This makes the solution for
the coefficients $c^{(L)}_{jk}$ ill-determined [see the general linear
least squares section in \cite{press} for a detailed discussion on
this issue]. Typically what is observed is that the inclusion of many
more low-variance EOFs only marginally improves the $r^{(L)}_k$
values, but that the corresponding patterns describe less variance and
become ``noisier'' i.e. project onto Z500 variations at progressively
smaller wavelengths. The optimal value of $L$ in a statistical
procedure like this, denoted by $L_c$, is subjective, but nevertheless
can be found from a tradeoff between the amount  of variance that the
patterns describe and their $r$-values.
\begin{figure*}
\begin{center}
\begin{minipage}{0.48\linewidth}
\includegraphics[width=0.64\linewidth,angle=270]{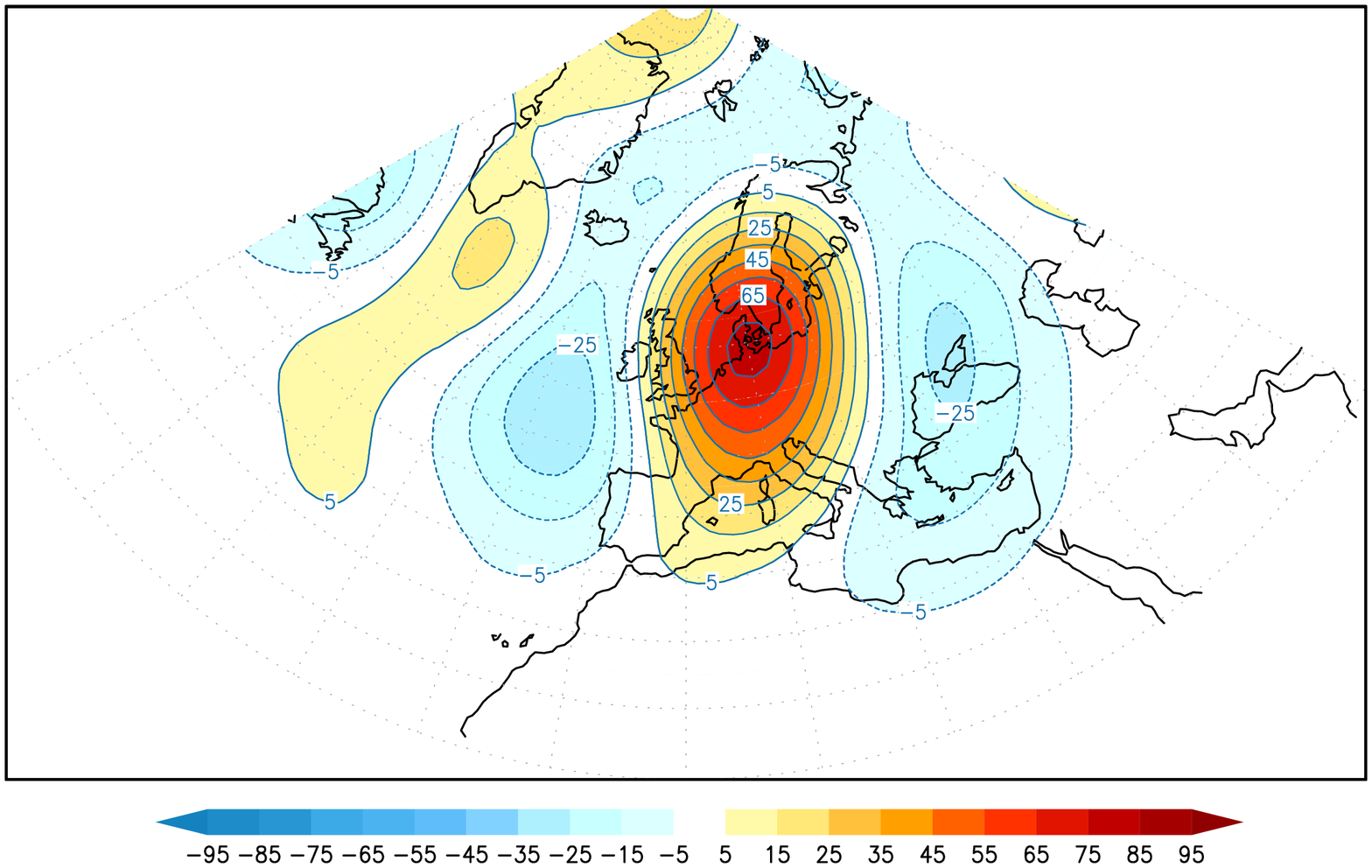}
\end{minipage}
\hspace{2mm}
\begin{minipage}{0.48\linewidth}
\includegraphics[width=0.64\linewidth,angle=270]{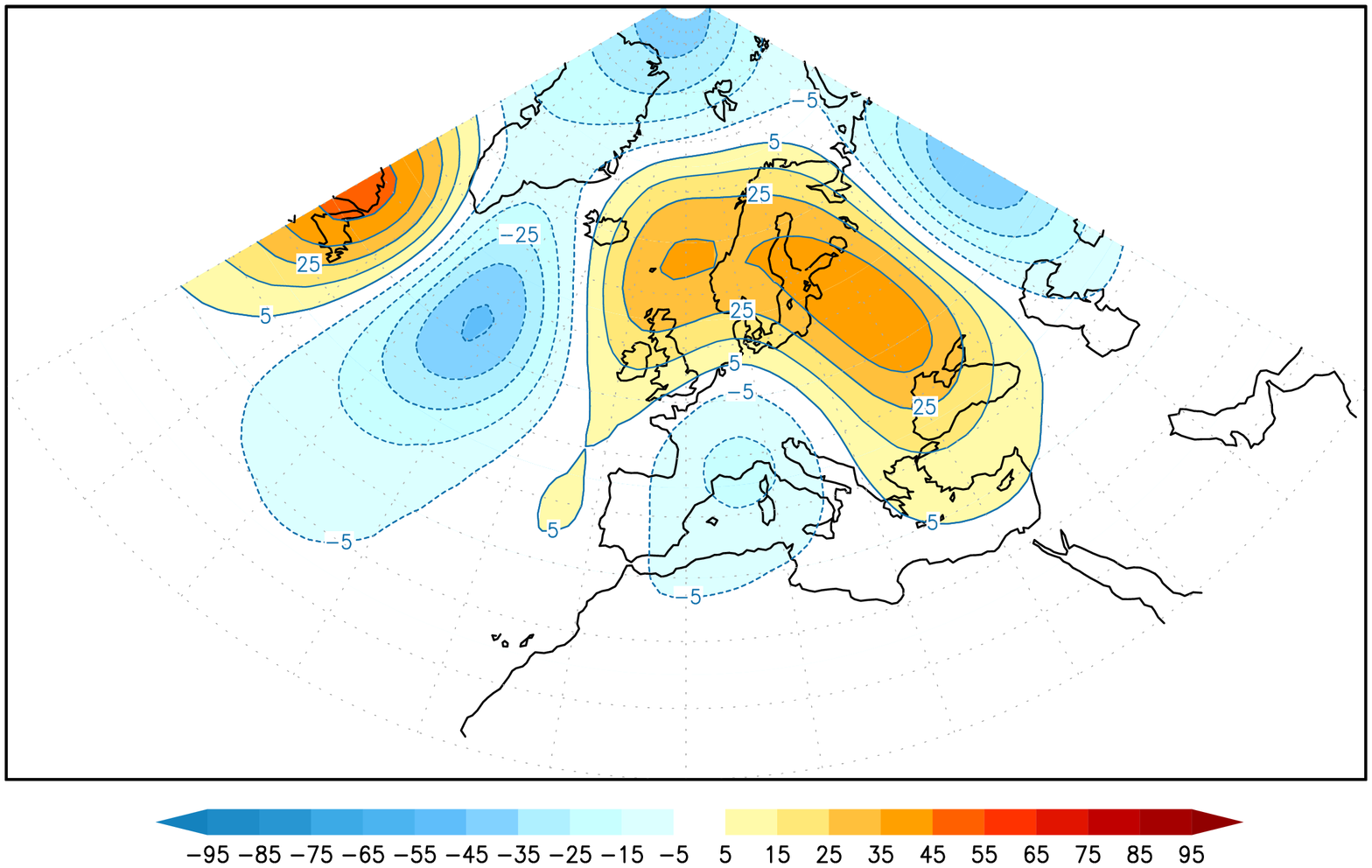}
\end{minipage}

\vspace{5mm}
\begin{minipage}{0.48\linewidth}
\includegraphics[width=0.64\linewidth,angle=270]{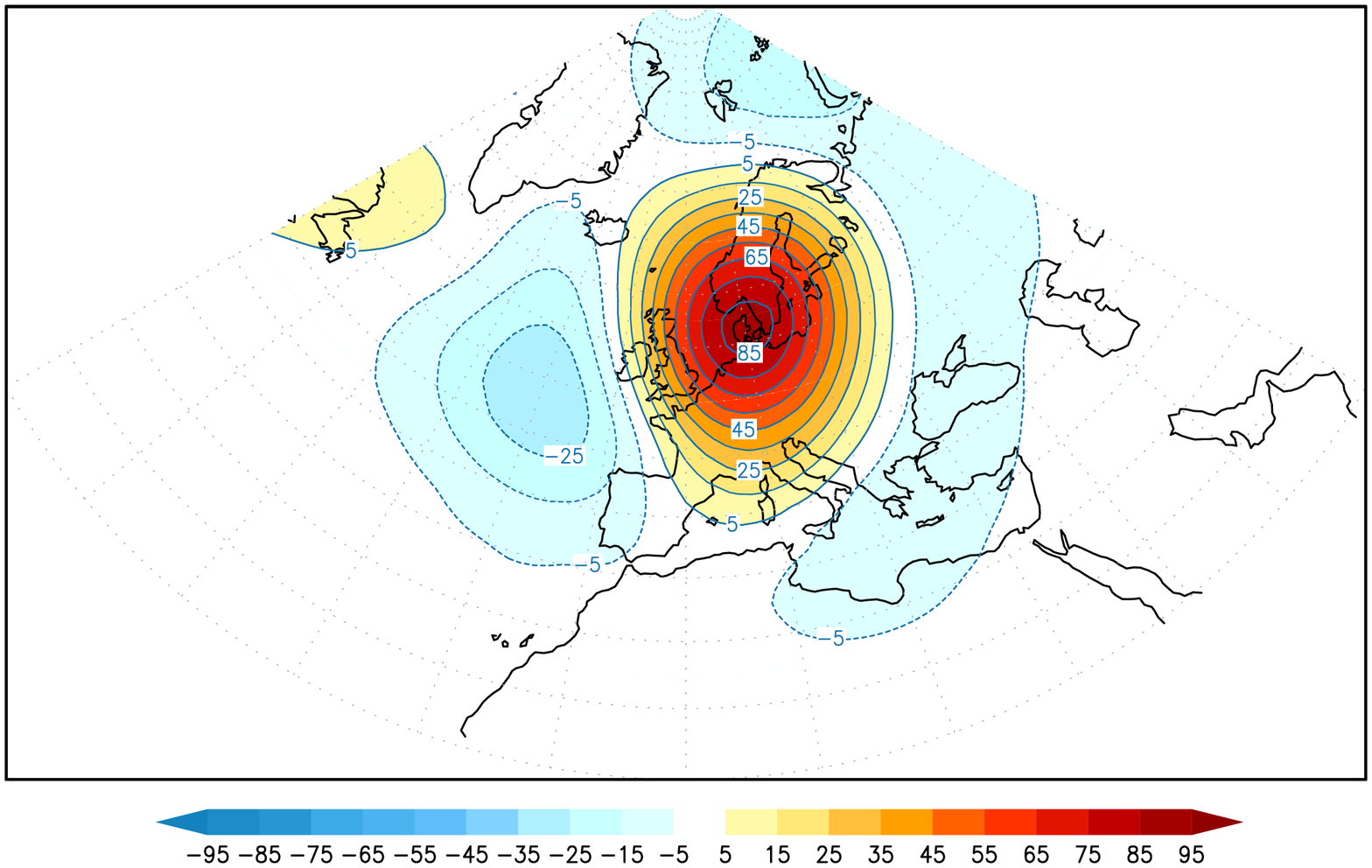}
\end{minipage}
\hspace{2mm}
\begin{minipage}{0.48\linewidth}
\includegraphics[width=0.64\linewidth,angle=270]{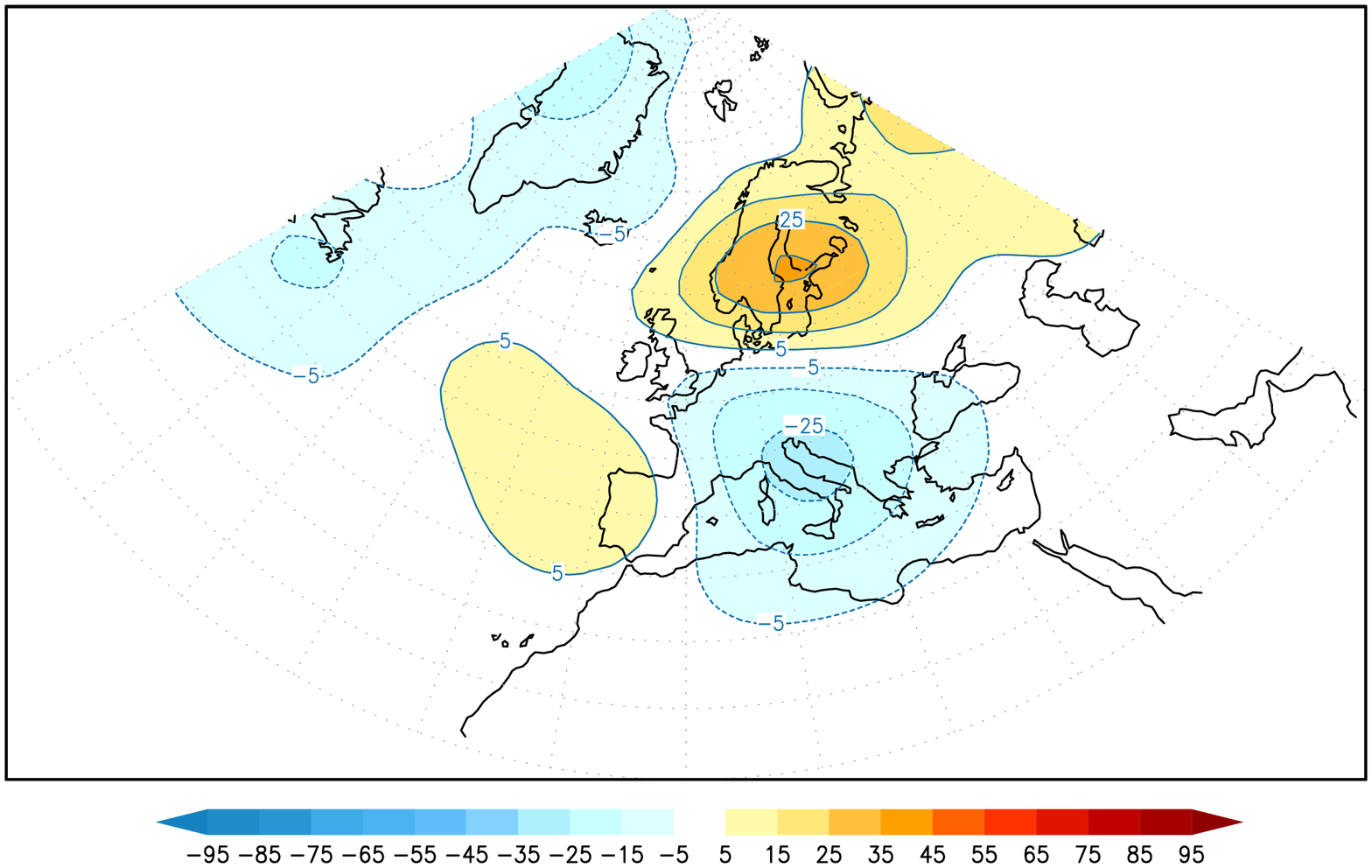}
\end{minipage}
\end{center}
\caption{The leading two Z500 daily anomaly patterns (EAFs) that are
associated with warm July and August daily temperatures in the
Netherlands: EAFs orthogonal in space, corresponding to $L_c=12$ (top
panel); EAF amplitudes uncorrelated in time, corresponding to
$L_c=50$ (bottom panel). The first EAFs are shown on the left, and the
second EAFs are shown on the right. All patterns have been multiplied
by one standard deviation of the corresponding amplitude time-series
(in meters).}
\label{fig5}
\end{figure*}

The procedure to determine $L_c$ for the daily summer (July and
August) temperature in the Netherlands [represented by T2m at
($52.5^\circ$N,$5^\circ$E)] and Z500 daily anomaly field over the
region $20^\circ$N-$90^\circ$N and $60^\circ$W-$60^\circ$E for 43
years (1958-2000) is as follows. As can be expected, both $r^{(L)}_1$
and $r^{(L)}_2$ are increasing functions with $L$
[Fig. \ref{fig4}(left)] and  the variance associated with the
corresponding EAFs tends to decrease with  increasing $L$ (not shown
here). For option (a), both $r^{(L)}_1$ and $r^{(L)}_2$ improve
significantly when including  EOF12 in the linear combination; at the
same time the variance of EAF1  decreases and the variance of EAF2
increases. Also the corresponding patterns change markedly. Between
$L=12$ and $L=15$ the patterns, $r$-values and variances remain
relatively unchanged. Beyond $L=15$ the $r$-values steadily increase,
the variance decreases and the patterns become
``noisier''. Simultaneously, the temporal correlation between the
dominant two EAF patterns steadily increases with $L$. For large $L$,
as Fig. \ref{fig4}(left) shows, both $r_1^{(L)}$ and $r_2^{(L)}$
values saturate to values very close to each other, and the solution
tends to become degenerate. Our interpretation of this is that the
information that is contained in the $Z500$ anomaly fields about the
local temperatures in the Netherlands is shared among increasingly
more patterns, which is an undesirable characteristic. For example,
for $L=12$, the temporal correlation between EAF1 and EAF2 is $0.58$,
for $L=50$ it is $0.93$. Based on these findings, we consider $L_c$ to
be equal to $12$.

A similar graph for EAFs calculated following option (b) are also
displayed in Fig. \ref{fig4}(right). By construction, the value of
$r^{(L)}_1$ is the same. In this case, the variance decreases as well
with increasing $L$, but much less so. The corresponding patterns are
quite stable beyond $L=19$. It is only beyond $L=200$ or so that the
second EAF more and more resembles the first EAF; for $L=19$ the
spatial correlation between EAF1 and EAF2 is only 0.2 (they are almost
orthogonal), for $L=200$ it is 0.4 and for $L=500$ it is 0.8. By
construction, the temporal correlation between EAF1 and EAF2 remains
zero. In this case, the choice of $L$ is not so critical and we simply
choose $L_c=50$.

The results for the spatially orthogonal EAFs corresponding to
$L_c=12$ and that for EAFs uncorrelated in time corresponding to
$L_c=50$ are shown in Fig. \ref{fig5}. The first EAFs obtained from
options (a) and (b) are very similar; the differences in the second
are bigger. The first corresponds to a high pressure system, leading
to clear skies over the Netherlands, an abundance of sunshine and a
warm southeasterly flow. In addition to this circulation anomaly, the
method finds another pattern that occurs less often; EAF2 corresponds
to an easterly flow regime bringing warm dry continental air masses to
the Netherlands. Option (b) gives a more localized $Z500$ anomaly
pattern, with a warm, easterly flow into the Netherlands. Option (a)
also captures the warm, easterly flow, but is less localized and is
less well defined as a function of $L$. The $r^{(L)}_2$ value is
larger for option (a), but it is temporally correlated to the first
EAF. This implies that part of the information about the local warm
temperatures in the Netherlands that is contained in the amplitude
timeseries of EAF2 is already captured by EAF1; they are not
independent. The $r^{(L)}_2$ value is smaller for option (b), but at
least the information it contains about the local warm temperatures in
the Netherlands is independent from EAF1. Given these considerations,
we conclude option (b), constructing EAFs that are temporally
uncorrelated is the best option.
\begin{figure*}
\begin{center}
\begin{minipage}{0.48\linewidth}
\includegraphics[width=0.8\linewidth,angle=270]{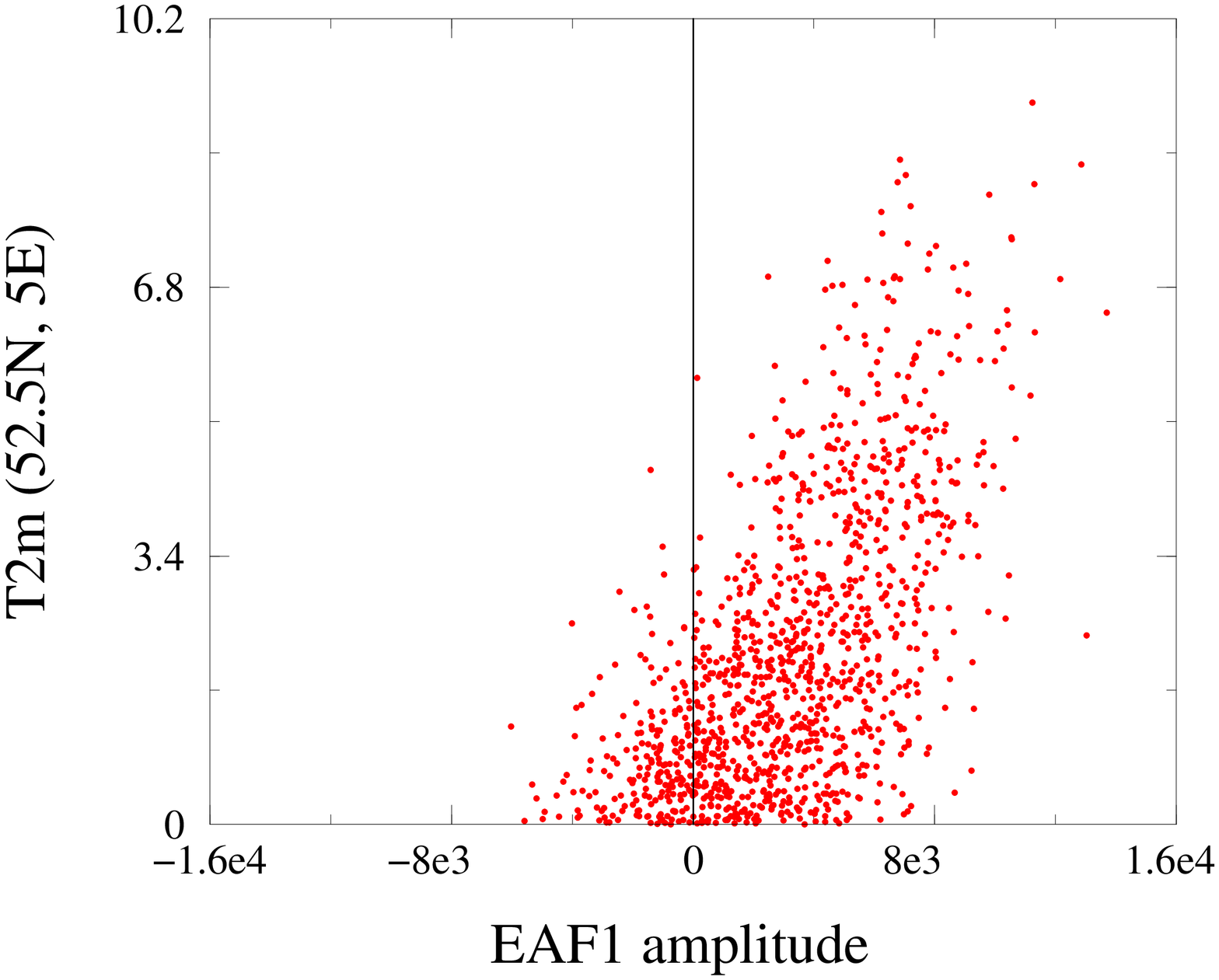}
\end{minipage}
\hspace{2mm}
\begin{minipage}{0.48\linewidth}
\includegraphics[width=0.8\linewidth,angle=270]{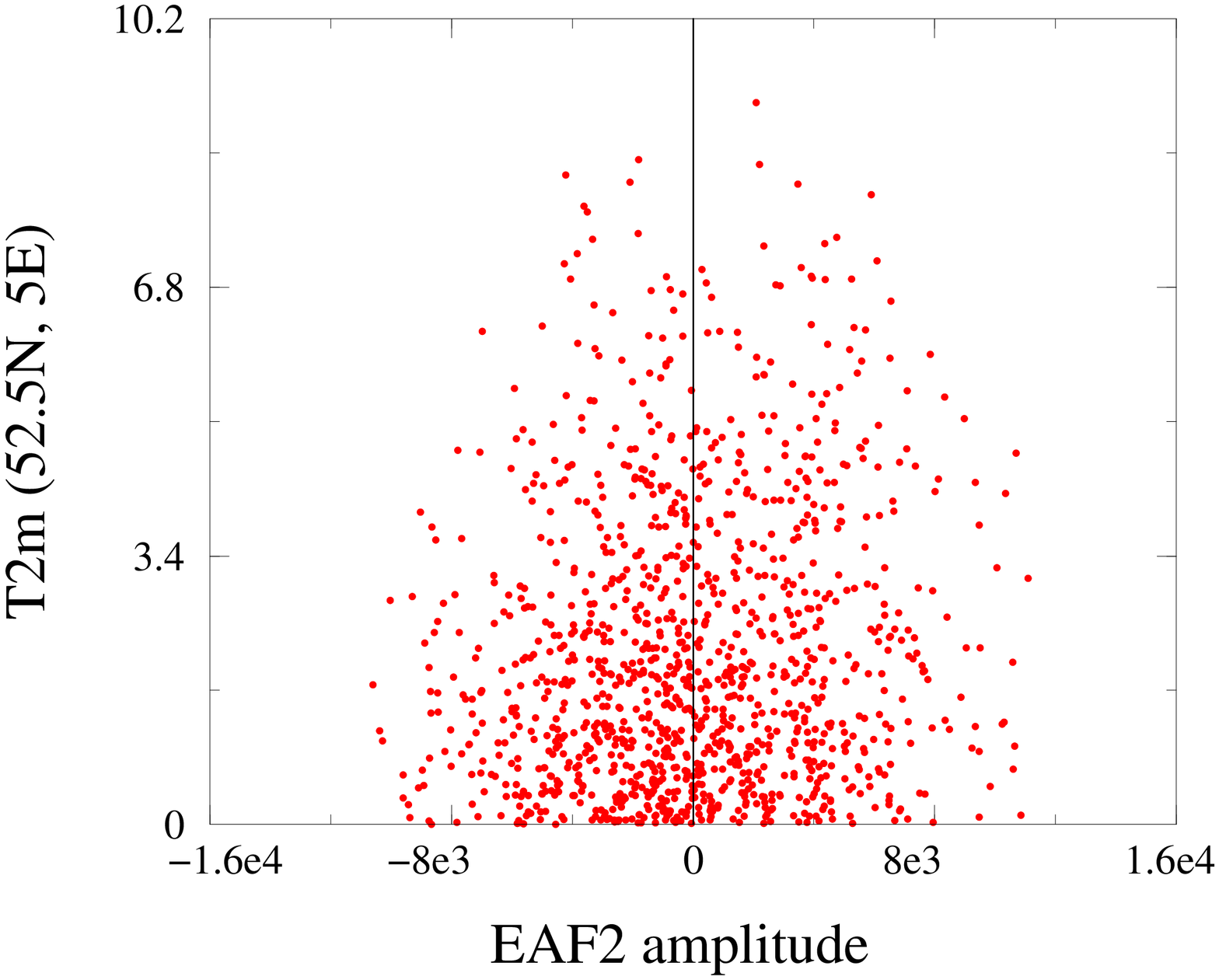}
\end{minipage}
\end{center}
\caption{Scatter plots for the amplitudes of EAF1 (left) and EAF2
(right) that are uncorrelated in time, corresponding to $L_c=50$,
against the daily mean two meter temperature in the Netherlands.}
\label{fig6}
\end{figure*}
\begin{figure*}
\begin{center}
\begin{minipage}{0.48\textwidth}
\includegraphics[width=0.64\linewidth,angle=270]{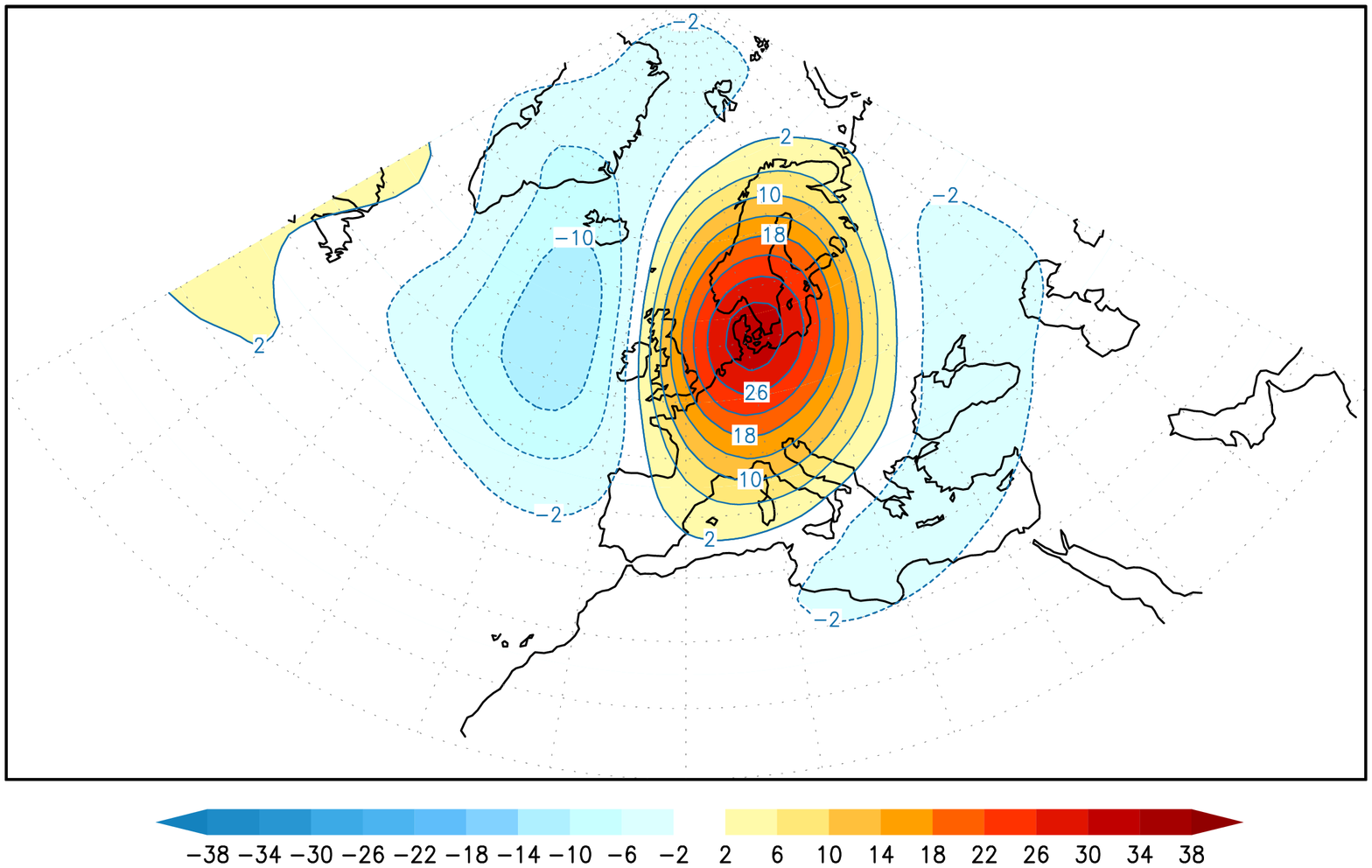}
\end{minipage}
\hspace{2mm}
\begin{minipage}{0.48\textwidth}
\includegraphics[width=0.64\linewidth,angle=270]{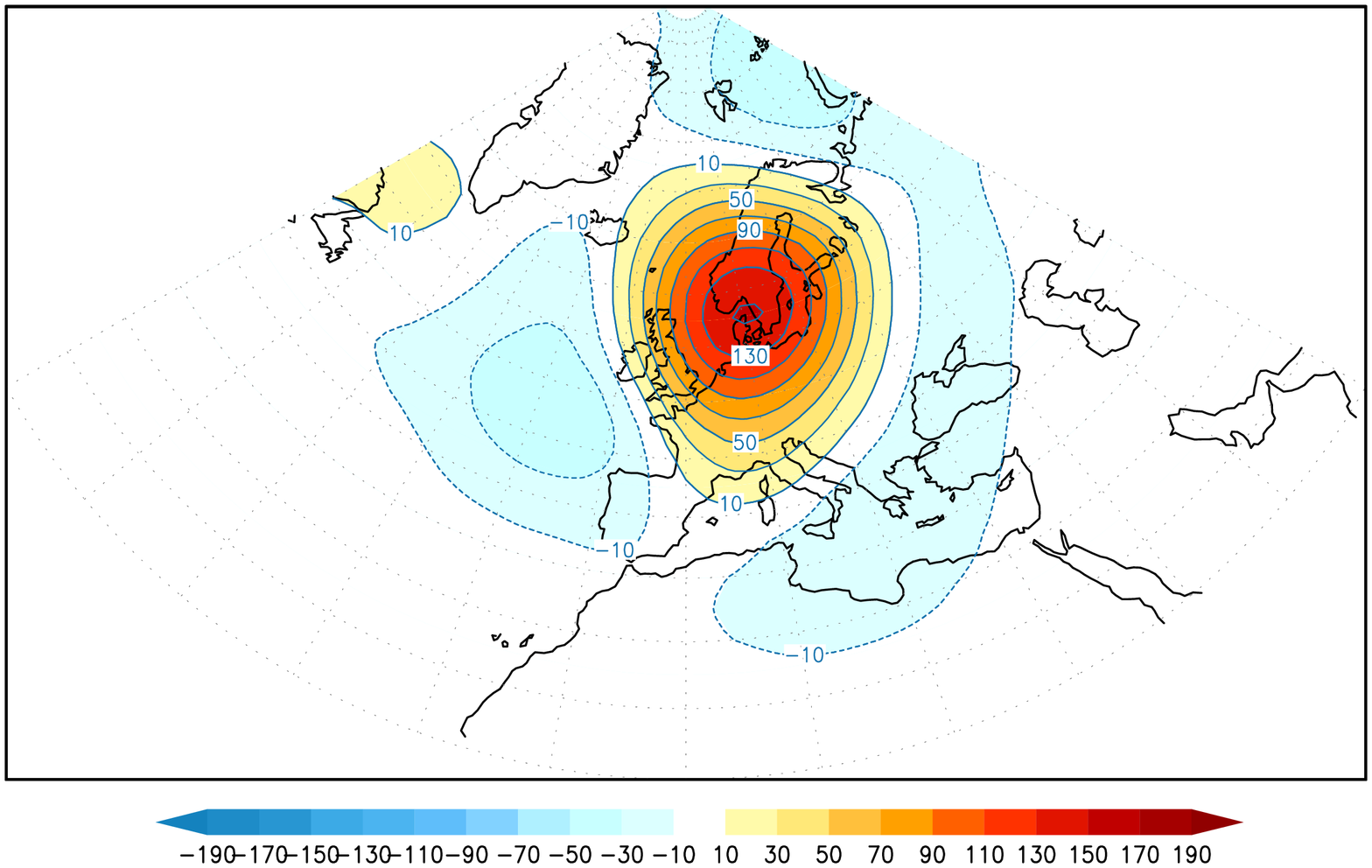}
\end{minipage}
\end{center}
\caption{Daily Z500 anomaly field regressed on daily mean temperatures
in The Netherlands in July and August in meters/Kelvin
(left). Composite Z500 daily anomaly field for 5\ warmest days in
July and August in the Netherlands over the period 1958-2000 in
meters. }
\label{fig7}
\end{figure*}

The scatterplots of $b^{(L_c=50)}_1$ and $b^{(L_c=50)}_2$ against the
positive temperature anomalies in the Netherlands for EAF1 and EAF2
that are uncorrelated in time are shown in Fig. \ref{fig6}. Compared
to the EOF with  the largest $r$ value (EOF1, see Fig. \ref{fig3}),
the relationship of $b^{(L_c=50)}_1$ to temperature is much stronger.
The $r$ value of the first EAF is almost a factor of 2 larger.  The
main contribution to the first EAF is from the first EOF, but also
EOFs 3,4 and 6 contribute substantially. Only two EAFs are found with
a clear connection (i.e., a tilt in the scatterplot) to warm extremes
in the Netherlands.  This information was spread mainly between EOFs
1, 3, 4 and 6.  Regressing Z500 anomalies upon the temperature
time-series in the Bilt gives a pattern that resembles EAF1
(Fig. \ref{fig7}). Also a simple compositing (averaging the 5 percent
hottest days) yields a pattern very similar to EAF1
(Fig. \ref{fig7}). In addition to this, the EAF  method is able to
identify another, less dominant, flow configuration that leads to warm
weather in the Netherlands through advection of warm airmasses from
eastern Europe. Comparing EAF1 to the clusters of summer Z500
anomalies published in \cite{cassou}, we note that EAF1 is a
combination of their `blocking' and `Atlantic low' regimes that favour
warm conditions in all of France and Belgium (temperatures in the
Netherlands were not analyzed). The easterly flow regime is not
present in their clusters.

In order to check that this method to identify the relevant
large-scale atmospheric circulation patterns for warm days in the
Netherlands is robust, we have also performed the same analysis for
the first 21 years (1958-1978) of daily summertime data and the last
21 years (1980-2000). In both cases we found very similar EAF1
patterns  and corresponding scatter plots as for the full period. EAF2
however is only recovered in the second period. One interpretation of
this is that EAF2 is less frequently present in the first period. As
argued by \cite{liu} this variation could be entirely due to the
chaotic nature of the atmospheric circulation and need not be caused
by a factor external to the atmosphere (as for instance increasing
levels of greenhouse gases, changes in sea surface temperatures or
solar activity to name a few).

Instead of taking all positive temperature anomalies, a threshold
could be introduced to Analise only the more extreme warm
days. However, limiting the analysis to  the 30\% warmest positive
temperature anomalies  did not qualitatively change the first two
EAFs. Also varying the value of the power applied to the temperature
anomaly from 1 to 3, only quantitatively modified the resulting EAFs,
but not qualitatively. A final test of robustness  was that we limited
the analysis to a smaller domain. Again we found the same  two EAF
patterns on a much smaller domain from 20 degrees east to 32.5 degrees
west and 35 to 70 degrees north. The method thus produces robust
patterns.

\conclusions[Discussion: Applicability of the Extreme Associated
  Functions\label{sec5}] 

The Extreme Associated Function method developed in this study to
establish the connection  between local weather extremes and
large-scale atmospheric circulation structures  has several
potentially useful applications.

First of all, since this method proved to satisfy several tests of
rigor and robustness for the temperature extremes in the Netherlands,
it can be applied for local temperature extremes at any other place,
or for that matter for other forms of extreme local weather conditions
as well, like precipitation or wind.  In this sense the method is
quite general.

EAFs can be used to evaluate the performance of climate models with
respect to the occurrence of local weather extremes. The EAF method
helps to answer the question whether the climate model is able to
generate the same patterns that are found in nature to be responsible
for local weather extremes with a similar probability of
occurrence in an objective manner. In addition, 
to evaluate the impact of climate change on local
weather extremes, the EAF method helps to answer the question whether
the probability of certain local weather extremes changes in
future scenario simulations due to a change in the probability of
occurrence of the EAFs.

It might be found that some climate models are able to simulate the
EAFs, but do not reproduce the local extremes well. \cite{lenderink}
for instance found that regional climate models forced with the right
large-scale circulation structures at the domain boundaries
nevertheless tended to overestimate the summer temperature
variability in Europe due to deficiencies in the description of the
hydrological cycle. The EAFs can be used to correct the model output
for this discrepancy by applying the observed statistical relationship
between the EAFs and the local extremes to the model generated EAFs.

By choosing the particular form of $r$ in Eq. (\ref{eq:r}) as the
quantity to be optimized, the EAF method turns out to be equivalent to
multiple linear regression. Other measures to describe the statistical
relationship between circulation and temperature present in the
scatterplot of Fig. \ref{fig3} could be designed that would make the
EAF method different from a multiple linear regression technique. In
this sense, the EAF method is more general and potentially can be
improved by designing a more apt measure.

\begin{acknowledgements}
We thank ECMWF for making the Z500 data 
publicly available. We also acknowledge the ENSEMBLES project, funded
by the European Commission's 6th Framework Programme through contract
GOCE-CT-2003-505539.
\end{acknowledgements}

\appendix
\section*{Calculation of the EAFs by a repetitive maximization
  procedure \label{append}} 

Since the entire appendix describes the procedure to calculate
$b^{(L)}_k$, i.e., the $b_k$-values for a given $L$, we drop all
superscripts involving $L$ for the sake of notational simplicity. 

\subsection*{A.1. Calculation of the first EAF\label{appenda}} 

\setcounter{equation}{0}
\renewcommand{\theequation}{A.1.\arabic{equation}}

To calculate which set of coefficients $c_{j1}$ maximize the value of
$r_1^2$ as expressed in  Eq. (\ref{eq:ralternative}) we take the
variation of $r_1^2$  w.r.t. variations $\delta c_{j1}$, and using
Eq. (\ref{eq:b}), obtain
\begin{eqnarray}
\delta r_1^2\,=\,2\times\nonumber\\&&
\hspace{-1.9cm}\frac{\displaystyle{\sum_{l=1}^{L}\!\!\left\{\!\langle
\tilde{T}(t)a_k(t)\rangle_{\mbox{\scriptsize p}}\langle \tilde{T}(t)a_l(t)\rangle_{\mbox{\scriptsize p}}\!-\!
[r_1^{\mbox{\scriptsize max}}]^2\!\langle
a_k(t)a_l(t)\rangle_{\mbox{\scriptsize p}}\!\right\}\!c_{l1}\delta c_{k1}}}{\langle
[b(t)]^2\rangle_{\mbox{\scriptsize p}}}\nonumber\\&&
\hspace{2.7cm}\quad\mbox{for}\,\,k=1,\ldots,L\,.
\label{eq:ea1}
\end{eqnarray}
This means that with the l.h.s. of Eq. (\ref{eq:ea1}) set to zero at
the maximum of $r_1^2$ for any choice of $\delta c_{k1}$, we obtain a
generalized eigenvalue equation: if we denote
$\langle\tilde{T}(t)a_k(t)\rangle_{\mbox{\scriptsize p}} \langle
\tilde{T}(t)a_l(t)\rangle_{\mbox{\scriptsize p}}$ by
$\tilde{a}_{k}\,\tilde{a}_{l}$, and $\left[\langle
a_k(t)a_l(t)\rangle_{\mbox{\scriptsize p}}\right]$ by $v^2_{kl}$ then
Eq. (\ref{eq:ea1}) leads to
\begin{eqnarray}
\sum_{l=1}^{L}\left\{\tilde{a}_k\,\tilde{a}_l\,-\,[r_1^{\mbox{\scriptsize max}}]^2\,
v^2_{kl}\right\}\,c_{l1}\,=\,0\,. 
\label{eq:ea2}
\end{eqnarray}

Equation (\ref{eq:ea2}) can be written as a matrix equation
\begin{eqnarray}
\mathbf{\tilde{A}}\,\mathbf{c}_1\,=[r_1^{\mbox{\scriptsize max}}]^2\,\mathbf{V}^2
\,\mathbf{c}_1, 
\label{eq:ea2mat}
\end{eqnarray}
where the $(k,l)$-th element of matrices $\mathbf{\tilde{A}}$ and
$\mathbf{V}^2$ are given by $\tilde{a}_k\,\tilde{a}_l$ and $v^2_{kl}$
respectively, and the $l$-th element of the column vector
$\mathbf{c}_1$ is given by $c_{l1}$. Note that in Eq. (\ref{eq:ea2})
$v^2_{kl}\not\propto\delta_{kl}$, since the time-average is defined
only over the days for which T2m$(t)\geq0$.  Since matrix
$\mathbf{\tilde{A}}$ is a tensor product of two column vectors
$\mathbf{\tilde{A}}=\mathbf{\tilde{a}}\mathbf{\tilde{a}}^{\mbox{\scriptsize T}}$,
where the superscript `T' indicates transpose, the matrix equation
(\ref{eq:ea2mat}) has only one eigenvector with non-zero eigenvalue,
given by
\begin{eqnarray}
\mathbf{c}_1 \propto \mathbf{V}^{-2}\mathbf{a}.
\end{eqnarray}
or equivalently
\begin{eqnarray}
c_{j1} \propto \sum_{i=1}^{L} \langle a_i(t)a_j(t) \rangle_{\mbox{\scriptsize p}}^{-1} \, \langle 
\tilde{T}_{\mbox{\scriptsize p}}(t)\,a_i(t) \rangle_{\mbox{\scriptsize p}}.
\label{eq:coeffmax}
\end{eqnarray}
The corresponding optimized value $r_1$ is determined from
Eq. (\ref{eq:ea2mat}).  

The equivalence between the maximization of $r_1^2$ and the multiple
linear regression of $\tilde{T}_{\mbox{\scriptsize p}}(t)$ on the timeseries of
the EOF amplitudes  $a_k(t)$'s [see Eq. (\ref{eq:regression})] is
apparent by noticing that the above solution for $\mathbf{c}_1$ is the
same as the solution of the multiple linear regression problem  given
in Eq. (\ref{eq:rcoeff}).

How the EAFs are determined from the coefficients $c_{jk}$ is shown in
the next section in which we explain the calculation of the remaining
$(L-1)$ EAFs.

\subsection*{A.2. Calculation of the remaining $(L-1)$ 
EAFs\label{appendb}} 

\setcounter{equation}{0}
\renewcommand{\theequation}{A.2.\arabic{equation}}

As explained in the text, the calculation of the remaining $(L-1)$
linear combinations requires a choice between two options. (a) The
patterns are orthogonal in space, or, (b) the amplitude timeseries are
uncorrelated in time. We will show the implementation of both options.

We first discuss option (a).

Combining the expansion of $\mathbf{Z500}(t)$ into EOFs
as in Eq. (\ref{eq:expansion}) and into EAFs as in
Eq. (\ref{eq:expansionf}) gives the following relation between the
EOFs and EAFs 
\begin{eqnarray}
\mathbf{e}_i = \sum_{k=1}^{L} c_{ik}\,\mathbf{f}_k \,\,\,\,\,\mbox{for}\,\,i=1,\ldots,L
\label{eq:releofeaf}
\end{eqnarray}

Option (a) demands the EAFs to be orthonormal in space which leads to
the following  condition for the corresponding coefficients $c_{jk}$
where we start from the  orthonormality condition of the EOFs
\begin{eqnarray}
\hspace{-4mm}\mathbf{e}_i \cdot \mathbf{e}_j=\sum_{k,l=1}^{L}
 c_{ik}\,c_{jl}\,\mathbf{f}_k \cdot \mathbf{f}_l =\sum_{k=1}^{L}
 c_{ik}\,c_{jk} = \delta_{ij}.
\end{eqnarray}
Additionally, it can be easily shown that
\begin{eqnarray}
\sum_{i=1}^{L} c_{ik}\,c_{il} = \delta_{kl}.
\label{eq:corthoa}
\end{eqnarray}
Using Eq. (\ref{eq:corthoa}), it is now straightforward to show from
Eq. (\ref{eq:releofeaf}) that the EAFs can be calculated from the EOFs
as
\begin{eqnarray}
\mathbf{f}_k = \sum_{i=1}^{L} c_{ik}\,\mathbf{e}_i \,\,\,\,\mbox{for}\,\,k=1,\ldots,L.
\label{eq:releafeofa}
\end{eqnarray}
Using this definition for the EAFs, the corresponding amplitudes
$b_k(t)$ are found by 
\begin{eqnarray} 
b_k(t)=\mathbf{f}_k \cdot \mathbf{Z500}(t).
\label{eq:bZ500a}
\end{eqnarray}

We now discuss option (b). 

For option (b), Eqs. (\ref{eq:expansionf}) and (\ref{eq:bZ500a})
cannot hold simultaneously. To obtain the coefficients $c_{ij}$ for
this option, we start with Eqs. (\ref{eq:expansionf}) and
define
\begin{eqnarray} 
b_k(t) = \sum_{i=1}^{L} c_{ik}\,a_i(t)= \sum_{i=1}^{L} c_{ik}\, \mathbf{e}_k \cdot \mathbf{Z500}(t) \nonumber\\&&
\hspace{-5.95cm} \equiv \mathbf{g}_k 
\cdot \mathbf{Z500}(t)
\label{definebi}
\end{eqnarray}
Then the conditions that $b_k(t)$ and $b_l(t)$ are uncorrelated in
time, i.e.,
\begin{eqnarray}
\langle b_k(t)\,b_l(t) \rangle =\delta_{kl} 
\label{uncorr}
\end{eqnarray}
yields, using the fact that the EOF amplitudes are uncorrelated in time,
\begin{eqnarray}
\sum_{i,j}^{L} c_{ik} \langle a_i(t)\,a_j(t) \rangle c_{jl} = \sum_{i}^{L} c_{ik}\, \sigma^2_i\, c_{il} = \delta_{kl}.
\label{eq:corthob}
\end{eqnarray}
We then define
\begin{eqnarray}
\mathbf{f}_k = \sum_{i=1}^{L} c_{ik}\,\sigma_i^2\,\mathbf{e}_i \,\,\,\,\mbox{for}\,\,k=1,
\ldots,L
\label{eq:releafeofb}
\end{eqnarray}
in terms of which Eq. (\ref{uncorr}) can be re-expressed as
\begin{eqnarray} 
\mathbf{f}_k \cdot \mathbf{g}_l =\delta_{kl}.
\label{eq:forthog}
\end{eqnarray}

Note here that for option (a) the patterns $\mathbf{f}_k$ are
automatically normalized to unity. For option (b), the patterns
$\mathbf{f}_k$ can be normalized to one, but the normalization of
$\mathbf{g}_k$ should be adjusted as well in order for
Eq. (\ref{eq:forthog}) to remain valid. 
 
To obtain the rest of the $(L-1)$ EAFs, 
the procedure described in Appendix A.1 needs to be repeated
$L-1$ times, but certain care needs to be taken because of the
orthonormality condition imposed by the definition of the set of
EAFs. When these subtle issues are taken into account, the 
procedure becomes a repetition of the following three steps.

\begin{itemize}
\item[(i)] Construct $\mathbf{Z500}'(t)$, the Z500 daily
anomaly field that lies within the vector subspace of the first $L$
EOFs but orthogonal to the first EAF. This is achieved in the
following manner. 

First define 
\begin{eqnarray}
\mathbf{e}'_j = \mathbf{e}_j-  (\mathbf{e}_j \cdot
\mathbf{f}_1)\,\mathbf{f}_1=\mathbf{e}_j - c_{j1}\,\mathbf{f}_1
\label{defineeprime}
\end{eqnarray}
for $j=2,\ldots,L$. The dot product of both sides of
Eq. (\ref{defineeprime}) with $\mathbf{Z500}(t)$ then yields
\begin{eqnarray}
\hspace{-2mm} a'_j(t) = a_j(t) - c_{j1}\,b_1(t)
\quad\mbox{for}\,\,j=2,\ldots,L
\label{eq:aprimea}
\end{eqnarray}
for option (a). For option (b), the corresponding expression is
\begin{eqnarray}
a'_j(t)\!= a_j(t)-c_{j1}\,\sigma_j^2\,b_1(t)\,\mbox{for}\,\,j=2,\ldots,L.
\label{eq:aprimeb}
\end{eqnarray}

\item[(ii)] Calculate the coefficients $c'_{j2}$ for $j=2,\ldots,L$
  that maximize $r_2$. 
\begin{eqnarray}
r_{2}c'_{j2} = \sum_{i=1}^{L} \langle a'_i(t)a'_j(t)\rangle_{\mbox{\scriptsize p}}^{-1}\,\langle \tilde{T}_{\mbox{\scriptsize p}}(t)\,a'_i(t) \rangle_{\mbox{\scriptsize p}}
\label{eq:r2coeff}
\end{eqnarray} 
with
\begin{eqnarray}
b_2(t)=\sum_{j=2}^{L} c'_{j2}\,a'_{j}(t)\equiv \sum_{j=1}^{L} c_{j2}\,a_{2}(t).
\label{eq:b2}
\end{eqnarray}

\item[(iii)] Next the coefficients $c_{j2}$ are calculated from the coefficients $ c'_{j2}$. For option (a) substitution of Eq. (\ref{eq:aprimea}) into Eq. (\ref{eq:b2}) leads to
\begin{eqnarray}
c_{j2} = c'_{j2} - c_{j1}\,\sum_{i=1}^{L}\,c'_{i2}\,c_{i1}\,\mbox{for}\,\,j=2,\ldots,L,
\end{eqnarray}
with the convention that $c'_{12}=0$. For option (b) substitution of
Eq. (\ref{eq:aprimeb}) into Eq. (\ref{eq:b2}) leads to
\begin{eqnarray}
c_{j2}\!= c'_{j2} - c_{j1}\sum_{i=1}^{L}\!c'_{i2}\,\sigma_i^2\,c_{i1}\,\mbox{for}\,\,j=2,\ldots,L,
\end{eqnarray}
with the convention that $c'_{12}=0$.

These steps are to be repeated until all $L$ coefficient vectors have been determined. For option (a) the EAFs are then determined from Eq. (\ref{eq:releafeofa}), for option (b) from Eq. (\ref{eq:releafeofb}).
\end{itemize}

\bibliographystyle{apalike}
\bibliography{atmos}

\begin{thebibliography}{}

\bibitem[Berner and Branstator, 2007]{berner}
Berner, J. and Branstator, G. (2007).
\newblock Linear and nonlinear signatures in the planetary wave dynamics of an
  agcm: probability densitiy functions.
\newblock {\em J. Atmos. Sci.}, 64:117--136.

\bibitem[Cassou et~al., 2005]{cassou}
Cassou, C., Terray, L., and Phillips, A. (2005).
\newblock Tropical {A}tlantic influence on {E}uropean heat waves.
\newblock {\em J. Climate}, 18:2805--2811.

\bibitem[Stephenson and O'Neill, 2004]{stephenson}
D.~Stephenson, A.~H. and O'Neill, A. (2004).
\newblock On the existence of multiple climate regimes.
\newblock {\em Quart. J. Roy. Meteor. Soc.}, 130:583--605.

\bibitem[Ferranti and Viterbo, 2006]{ferranti}
Ferranti, L. and Viterbo, P. (2006).
\newblock The {E}uropean summer of 2003: sensitivity to soil water initial
  conditions.
\newblock {\em J. Climate}, 19:3659--3680.

\bibitem[Greatbatch and Rong, 2006]{greatbatch}
Greatbatch, R. and Rong, P.~P. (2006).
\newblock Discrepancies between different northern hemisphere summer
  atmospheric data products.
\newblock {\em J. Climate}, 19:1261--1273.

\bibitem[Hsu and Zwiers, 2001]{hsu}
Hsu, C.~J. and Zwiers, F. (2001).
\newblock Climate change in recurrent regimes and modes of northern hemisphere
  atmospheric variability.
\newblock {\em J. Geophys. Res.}, 106:20145--20159.

\bibitem[Kysely, 2002]{kysely}
Kysely, J. (2002).
\newblock Temporal fluctuations in heat waves at {P}rague-{K}lementinum, the
  {C}zech {R}epublic, from 1901-97, and their realationships to atmospheric
  circulation.
\newblock {\em Int. J. Climatol.}, 22:33--50.

\bibitem[Lenderink et~al., 2007]{lenderink}
Lenderink, G. et~al. (2007).
\newblock Summertime inter-annual temperature variability in an ensemble of
  regional model simulations:analysis of the surface energy budget.
\newblock {\em Climatic Change}, 81:233--247.

\bibitem[Liu and Opsteegh, 1995]{liu}
Liu, Q. and Opsteegh, J. (1995).
\newblock Interannual and decadal varations of blocking activity in a
  quasi-geostrophic model.
\newblock {\em Tellus}, 47A:941--954.

\bibitem[Malone et~al., 1984]{malone}
Malone, R. et~al. (1984).
\newblock The simulation of stationary and transient geopotential-height eddies
  in {J}anuary and {J}uly with a spectral general circulation model.
\newblock {\em J. Atmos. Sci.}, 41:1394--1419.

\bibitem[North et~al., 1982]{north}
North, G. et~al. (1982).
\newblock Sampling errors in the estimation of empirical orthogonal functions.
\newblock {\em Mon. Weather Rev.}, 110:699--706.

\bibitem[Pelly and Hoskins, 2003]{pelly}
Pelly, J.~L. and Hoskins, B.~J. (2003).
\newblock How well does the {ECMWF} ensemble prediction system predict
  blocking?
\newblock {\em Q. J. R. Meteorol. Soc.}, 129:1683--1702.

\bibitem[Plaut and Simonnet, 2001]{plaut}
Plaut, G. and Simonnet, E. (2001).
\newblock Large-scale circulation classification weather regimes, and local
  lcimate over {F}rance, the {A}lps and {W}estern {E}urope.
\newblock {\em Clim. Res.}, 17:303--324.

\bibitem[Press et~al., 1986]{press}
Press, W.~H., B.~P. Flannery, S.~A. Teukolsky and W.~T. Vetterling, (1986).
\newblock Numerical Recipes.
\newblock {\em Cambridge University Press}, isbn 0521308119, 818pp.

\bibitem[Sanchez-Gomez and Terray, 2005]{sanchez}
Sanchez-Gomez, E. and Terray, L. (2005).
\newblock Large-scale atmospheric dynamics and local intense precipitation
  episodes.
\newblock {\em Geophys. Res. Lett.}, 32:L2471.

\bibitem[Schaeffer et~al., 2005]{schaef}
Schaeffer, M., Selten, F.~M., and Opsteegh, J.~D. (2005).
\newblock Shifts of means are not a proxy for changes in extreme winter
  temperatures in climate projections.
\newblock {\em Clim. Dyn.}, 25:51--63.

\bibitem[Sch{\"a}r et~al., 2004]{scharetal}
Sch{\"a}r, C. et~al. (2004).
\newblock The role of increasing temperature variability in {E}uropean summer
  heat waves.
\newblock {\em Nature}, 427:332--336.

\bibitem[Selten et~al., 1999]{selten}
Selten, F., Haarsma, R., and Opsteegh, J. (1999).
\newblock On the mechanism of north {A}tlantic decadal variability.
\newblock {\em J. Climate}, 12:1256--1973.

\bibitem[van Ulden and van Oldenborgh, 2006]{ulden}
van Ulden, A. and van Oldenborgh, G. (2006).
\newblock Large-scale atmospheric circulation biases and changes in global
  climate model simulations and their importance for climate change in
  {C}entral {E}urope.
\newblock {\em Atmos. Chem. Phys.}, 6:863--881.

\end{thebibliography}

\end{document}